\documentclass[twocolumn]{aastex701}

\graphicspath{{./}{figures/}}

\usepackage{amsmath}

\begin{document}

\title{Recombination Clumping Factor of Physically Defined Intergalactic Medium at the Epoch of Reionization}

\correspondingauthor{Yuri Oku, Renyue Cen}
\author[orcid=0000-0002-5712-6865,gname='Yuri',sname='Oku']{Yuri Oku}
\affiliation{Center for Cosmology and Computational Astrophysics, Institute for Advanced Study in Physics
Zhejiang University, Hangzhou 310058, China}
\affiliation{Institute of Astronomy, School of Physics, Zhejiang University, Hangzhou 310058, China}
\affiliation{Theoretical Astrophysics, Department of Earth \& Space Science, Graduate School of Science, The University of Osaka, 1-1 Machikaneyama, Toyonaka, Osaka 560-0043, Japan}
\email[show]{yuri.oku.astro@gmail.com}

\author[orcid=0000-0001-8531-9536,gname=Renyue, sname='Cen']{Renyue Cen}
\affiliation{Center for Cosmology and Computational Astrophysics, Institute for Advanced Study in Physics
Zhejiang University, Hangzhou 310058, China}
\affiliation{Institute of Astronomy, School of Physics, Zhejiang University, Hangzhou 310058, China}
\email[show]{renyuecen@zju.edu.cn}

\begin{abstract}

The recombination clumping factor, $C$, is a key parameter in modeling cosmic reionization, but its value is sensitive to the definition of the Intergalactic Medium (IGM). We investigate the clumping factor using the \textsc{Gamer-2} adaptive mesh refinement cosmological hydrodynamical simulation code. We introduce a new, physically-motivated definition of the IGM based on the effective transmission factor of ionizing photons. 
We perform large-scale simulations with varying intensities of the uniform ultraviolet background, and we find that our physically-defined clumping factor is slightly lower than, yet comparable to, the values derived from traditional overdensity thresholds, within a factor of two. At $z=6$, we obtain a clumping factor of $C \sim 3$, consistent with previous studies, indicting that the clumping factor is robust to numerical resolution, box size, and the definition of the IGM.
Our zoom-in simulations further show that supernova feedback has two competing effects on reionization; it enhances recombination by increasing the density of ionized gas, while facilitating ionization by heating gas and reducing the neutral fraction. However, these effects are limited to the scales of $\sim$ 100 kpc and do not significantly alter the global clumping factor.

\end{abstract}

\keywords{\uat{Intergalactic medium}{813} --- \uat{Reionization}{1383} --- \uat{Hydrodynamical simulations}{767}}

\section{Introduction}

The first galaxies in the universe emitted ionizing photons, which transformed the intergalactic medium (IGM) from a neutral state to an ionized state, a process known as cosmic reionization \citep[for recent reviews, see][]{2022ARA&A..60..121R, 2022LRCA....8....3G}.
The evolution of the IGM's ionization state is governed by the balance between the rate at which ionizing photons are produced and the rate at which ionized hydrogen recombines.
The classical model introduced by \citet{1999ApJ...514..648M} describes the filling factor of ionized hydrogen, $Q_{\rm HII}$, as
\begin{equation}
    \frac{dQ_{\rm HII}}{dt} = \frac{\dot{n}_{\rm ion}}{\langle{n}_{\rm H}\rangle} - \frac{Q_{\rm HII}}{\bar{t}_{\rm rec}},
    \label{eq:madau_model}
\end{equation}
where $\dot{n}_{\rm ion}$ is the production rate of ionizing photons per unit volume, $\langle{n}_{\rm H}\rangle$ is the mean hydrogen number density, and $\bar{t}_{\rm rec}$ is the volume-averaged recombination time.

Because the recombination rate depends on the square of the local ionized gas density, it is highly sensitive to the small-scale structure, or ``clumpiness," of the IGM.
This clumpiness is commonly quantified by the clumping factor, defined as the ratio of the mean of the squared density to the square of the mean density, i.e., $C \equiv \langle{n^2}\rangle / \langle{n}\rangle^2$ \citep{1997ApJ...486..581G}.
The clumping factor has been refined over the decades, using simulations with higher resolution and more sophisticated physics, such as radiative transfer, as well as improving the formalism of the clumping factor by accounting for the spatial variation of the temperature \citep{2007ApJ...671....1T,2007ApJ...657...15K,2009MNRAS.394.1812P,2011ApJ...743...82M,2012MNRAS.427.2464F,2012ApJ...747..100S,2013ApJ...763..146E,2014ApJ...789..149S,2015ApJ...810..154K,2016ApJ...831...86P,2020ApJ...905..132C,2020ApJ...898..149D,2022MNRAS.511.4005K,2024MNRAS.528.1296C}.

However, the clumping factor can remain sensitive to the adopted definition of the IGM. 
Since the clumping factor is intended to quantify recombination in the diffuse IGM, it is essential to exclude regions associated with galaxies. 
This avoids double-counting recombinations that are already accounted for in the escape fraction of ionizing photons from galaxies.

Another important aspect is the definition of the `$Q$' in Eq.~(\ref{eq:madau_model}).
$Q$ is often interpreted as the volume fraction of ionized regions; however, the ionization and recombination rates account for the number of ionized hydrogen atoms in the IGM, which translates to the mass-weighted ionization fraction rather than the volume fraction.
\citet{2020ApJ...905..132C} proposed a mass-weighted ionization fraction model, which is more physically consistent with the ionization and recombination rates.
They also introduced a clumping factor definition that is consistent with the mass-weighted ionization fraction model, and demonstrated its effectiveness in reproducing the ionization history in their SCORCH radiation-hydrodynamic simulations.

In this article, we present a physically motivated definition of the IGM based on the effective optical depth to ionizing photons from external sources, and we re-formulate the recombination clumping factor accordingly.
Our cosmological hydrodynamical simulations adopt a uniform ultraviolet background (UVB) as the ionizing source, allowing us to study the reionization history and clumping factor of the IGM irradiated by external sources with high spatial resolution in a comoving simulation volume of $(20\,{\rm Mpc}/h)^3$.

The remainder of this article is organized as follows. In Section~\ref{sec:definition}, we present the physical definition of the IGM and the recombination clumping factor.
In Section~\ref{sec:fullbox}, we measure the clumping factor in large-volume cosmological simulations with different UVB intensities.
In Section~\ref{sec:zoom}, we measure the clumping factor in zoom-in simulations that resolve small-scale structures in the IGM and discuss the feedback effects on the clumping factor.
Finally, we summarize our findings and discuss their implications in Section~\ref{sec:conclusion}.

\section{Recombination Clumping Factor}
\label{sec:definition}

The ionization balance of the diffuse IGM in a comoving volume of $V$ is described by
\begin{equation}
    \frac{dN_{\rm HII}^{\rm IGM}}{dt} = \dot{N}_{\rm ion}^{\rm IGM} + \int_{\rm IGM}[\gamma_{\rm coll}n_{\rm HI}n_{\rm e} - \alpha n_{\rm HII}n_{\rm e}]dv,
\end{equation}
where $N_{\rm HII}^{\rm IGM}$ is the number of ionized hydrogen atoms in the IGM, $\dot{N}_{\rm ion}$ is the total photoionization rate in the comoving volume $V$, $\gamma_{\rm coll}$ is the collisional ionization coefficient, $\alpha $ is the recombination coefficient, and $n_{\rm HII}$, $n_{\rm HI}$, and $n_{\rm e}$ are the densities of ionized hydrogen, neutral hydrogen, and electrons, respectively.

Here, we define the diffuse IGM as the region traversed by the ionizing photons, i.e., not self-shielded by the neutral hydrogen. We also ignore collisional ionization, which is subdominant compared to photoionization in the low-density IGM.
Then, the integral over the IGM in the equation above can be rewritten as
\begin{equation}
    \frac{dN_{\rm HII}^V}{dt} =  \dot{N}_{\rm ion}^{\rm IGM}  - \int_{V} \alpha n_{\rm HII} n_{\rm e} \mathcal{F} \,dv,
    \label{eq2}
\end{equation}
where $\mathcal{F} \equiv e^{-\tau_{\rm eff}}$ is the transmission factor of the diffuse IGM, and $\tau_{\rm eff}$ is the effective optical depth to ionizing photons from external sources.
We have replaced $N_{\rm HII}^{\rm IGM}$ with $N_{\rm HII}^V$, the number of ionized hydrogen atoms in the comoving volume $V$, assuming that most of the ionized hydrogen resides in the IGM.

We obtain the time evolution of the mass-weighted ionization fraction, $\langle{ x_{\rm HII} }\rangle_{M}$, by dividing both sides of Eq.~(\ref{eq2}) by $V$ and the mean hydrogen number density $\langle{n}_{\rm H}\rangle_{V}$,
\begin{equation}
    \frac{  d\langle{ x_{\rm HII}  }\rangle_{M} }
         {  dt  } 
    = \frac{\langle \dot{n}_{\rm ion} \rangle_{V}}{\langle{n}_{\rm H}\rangle_{V}}
    \,-\, C_{{\rm rec}, \mathcal{F}}
           {  
              \langle{ \alpha            }\rangle_{V}
              \langle{ x_{\rm HII}       }\rangle_{M}
              \langle{ n_{\rm e}         }\rangle_{V}
              \langle{ \mathcal{F}       }\rangle_{V}
           },
    \label{eq3}
\end{equation}
where
\begin{equation}
    C_{{\rm rec}, \mathcal{F}} \equiv
    \frac{
    \langle{  \alpha n_{\rm HII}n_{\rm e}\mathcal{F}  }\rangle_{V}
    }
    {  
    \langle{ \alpha            }\rangle_{V}
    \langle{ n_{\rm HII}       }\rangle_{V}
    \langle{ n_{\rm e}         }\rangle_{V}
    \langle{ \mathcal{F}       }\rangle_{V}
    }
    \label{eq:crec}
\end{equation}
is the recombination clumping factor.
We have used notation for the volumetric average, $\langle{X}\rangle_V = \int_V X dv / V$, and the mass-weighted average, $\langle{X}\rangle_M = \int_V \rho_{\rm H} X dv / \int_V \rho_{\rm H} dv$, where $\rho_{\rm H}$ is the mass density of hydrogen.
We note the relation $\langle{ x_{\rm HII} }\rangle_{M} = \langle{ n_{\rm HII} }\rangle_{V} / \langle{ n_{\rm H} }\rangle_{V}$.

It is helpful to normalize the clumping factor to a fixed temperature and remove the $\langle{ \mathcal{F} }\rangle_{V}$ term in the denominator for practical use. Then, Eq.~(\ref{eq3}) becomes

\begin{equation}
    \frac{  d\langle{ x_{\rm HII}  }\rangle_{M} }
         {  dt  } 
    = \frac{\langle \dot{n}_{\rm ion} \rangle_{V}}{\langle{n}_{\rm H}\rangle_{V}}
    \,-\, C_{{\rm norm}, \mathcal{F}}
           {  
              \alpha (10^4\, {\rm K})
              \langle{ x_{\rm HII}       }\rangle_{M}
              \langle{ n_{\rm e}         }\rangle_{V}
           }
    \label{eq3_norm}
\end{equation}
with
\begin{equation}
    C_{{\rm norm}, \mathcal{F}} \equiv
    \frac{
    \langle{  \alpha n_{\rm HII}n_{\rm e}\mathcal{F}  }\rangle_{V}
    }
    {  
              \alpha (10^4\, {\rm K})
              \langle{ n_{\rm HII}       }\rangle_{V}
              \langle{ n_{\rm e}         }\rangle_{V}
    }
    \label{eq:cnorm}
\end{equation}
being the normalized clumping factor.

In Sections~\ref{sec:fullbox} and \ref{sec:zoom}, we evaluate the clumping factor using cosmological hydrodynamical simulations.
We adopt the fitting function for the case B recombination coefficient presented by \citet{1997MNRAS.292...27H}, taking into account that the mean free path of ionizing photons is shorter than the Hubble length for $z>6$ \citep{2024ApJ...971...75M}.
The effective optical depth is evaluated as $\tau_{\rm eff} = \frac{1}{2} \bar{\sigma}_{\rm ion} n_{\rm HI}^2 / |\nabla n_{\rm HI}|$ using the gray (spectrum-averaged) ionization cross-section for hydrogen,
$\bar{\sigma}_{\rm ion}\equiv
{
\int_{\nu_{\rm {HI}}}^{\nu_{\rm {HeII}}} \frac{J_\nu}{\nu} \sigma_{\rm {H}, \nu} d\nu
} / {
\int_{\nu_{\rm {HI}}}^{\nu_{\rm {HeII}}} \frac{J_\nu}{\nu} d\nu
}$,
where the $\nu_{\rm {HI}}$ and $\nu_{\rm {HeII}}$ are the frequencies at the hydrogen Lyman edge and \ion{He}{2} ionization edge, respectively. The $J_\nu$ is the UVB intensity, and $\sigma_{\rm {H}, \nu}$ is the photoionization cross-section for hydrogen \citep{2013MNRAS.430.2427R}.

\begin{figure}
    \centering
    \includegraphics[width=\linewidth]{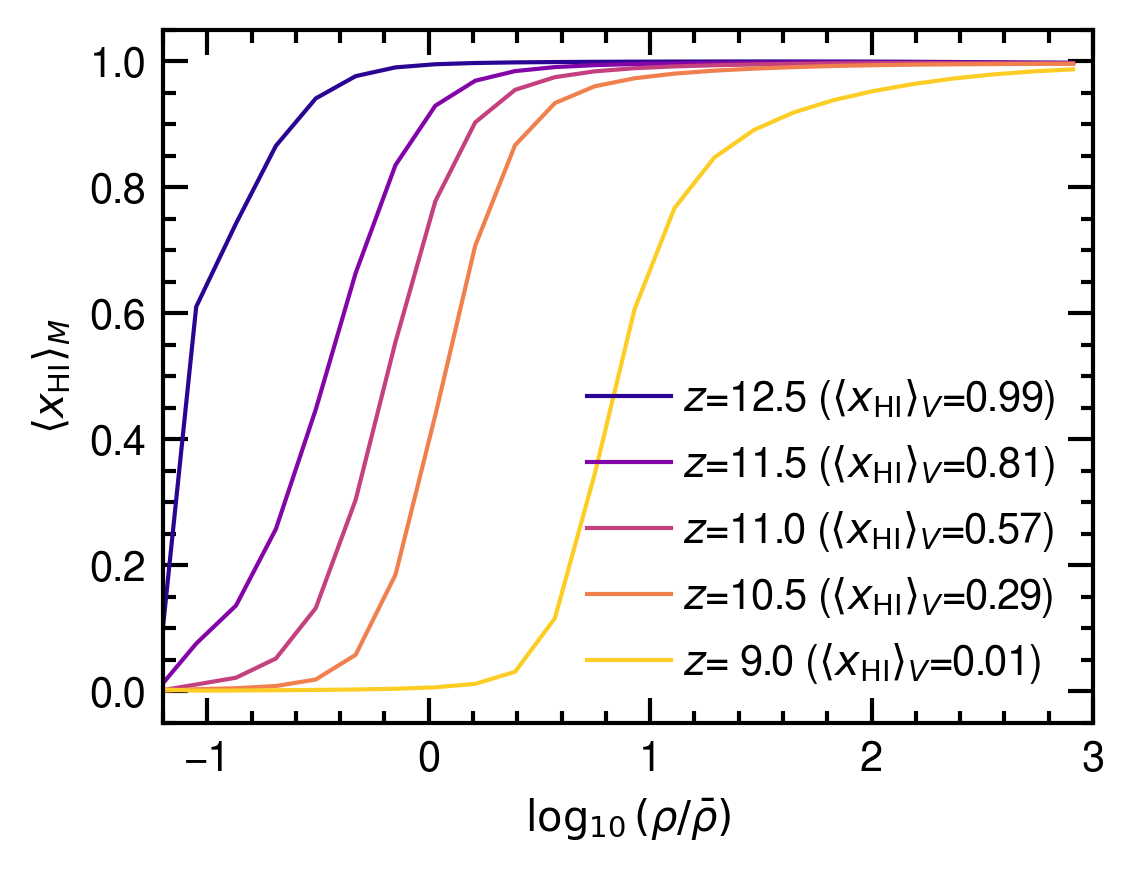}
    \caption{shows mass-weighted neutral fraction of hydrogen as a function of gas overdensity from the \texttt{HM12UVB} run at $z=$12.5, 11.5, 11.0, 10.5, and 9.0. The corresponding volume-weighted average of neutral fractions of hydrogen at each redshift are indicated in the legend.}
    \label{fig:xHIm_vs_density}
\end{figure}

The expression for $\tau_{\rm eff}$ is derived with assuming that the density gradient is linear. We have confirmed that the scale height, $n_{\rm HI}/|\nabla n_{\rm HI}|$, is resolved by [1, 2, 4, 8] cells in [97\%, 87\%, 60\%, 20\%] of the volume in the zoom-in simulation.
Figure~\ref{fig:xHIm_vs_density} shows the mass-weighted neutral fraction of hydrogen, $\langle x_{\rm HI} \rangle_M$, as a function of the gas overdensity, $\Delta \equiv \rho_{\rm gas}/\langle \rho_{\rm gas} \rangle_V$, from the \texttt{HM12UVB} run at different redshifts as indicated in the legend. The distribution functions of the neutral fraction show drop-off toward low density, which is qualitatively consistent with previous works that use radiative transfer to explore self-shielding effects \citep{2011ApJ...743...82M,2013ApJ...763..146E,2016ApJ...831...86P,2020ApJ...898..149D,2024MNRAS.528.1296C}.

\section{Full-Box Simulation}
\label{sec:fullbox}

\subsection{Method}
We generate the initial condition at $z=99$ using the cosmological initial condition generator \textsc{Music v1.53}\footnote{The code's website is \url{https://www-n.oca.eu/ohahn/MUSIC/index.html}} \citep{2011MNRAS.415.2101H}.
We assume a flat $\Lambda$ CDM cosmology with cosmological parameters $(\Omega_{\rm m}, \Omega_{\rm \Lambda}, \Omega_{\rm b}, \sigma_8, n_{\rm s}, h)=(0.3147, 0.6853, 0.0492, 0.8101, 0.9652, 0.6737)$, consistent with the Planck 2018 results \citep{2020A&A...641A...6P}.
The comoving volume of $(20\,h^{-1}\,{\rm Mpc})^3$ is covered by $1024^3$ uniform-resolution grids; the cell size at root level is 29.0 comoving kpc, the particle mass of dark matter is $m_{\rm DM} = 8.15\times10^5\, M_\odot$, and the mean mass of the gas cell is $m_{\rm gas, IC}=1.51\times10^5\, M_\odot$.

We perform cosmological hydrodynamical simulations using the adaptive mesh refinement (AMR) code \textsc{Gamer v2.1.1}\footnote{The code's website is \url{https://github.com/gamer-project/gamer}} \citep{2018MNRAS.481.4815S} with the MUSCL-Hancock scheme, the HLLC Riemann solver, and the piecewise linear reconstruction method.
The self-gravity of dark matter and baryons is computed by solving the Poisson equation using the successive-over-relaxation method with the fast Fourier transform method at the base level.

The mesh refinement is allowed to an additional four refinement levels to the root level, and the finest cell size is 1.81 comoving kpc, which is enough to resolve the Jeans length of the cosmic filaments in the IGM.
\textsc{Gamer-2} adopts the so-called block-based AMR structure, and we refine a patch (block) consisting of $8^3$ cells into eight patches when either of the following refinement criteria is satisfied: 1) the number of particles in the patch, including dark matter and stars, exceeds 2048, 2) the gas mass of any cell in the patch exceeds $4\,m_{\rm gas,IC}$, or 3) the cell length exceeds one fourth of the local Jeans length of any cell in the patch.
The Jeans refinement criterion is applied only to the refinement of levels 2 and 3.

For the treatment of radiative cooling and star formation, we follow the common physics model of the \texttt{CosmoRun} of the AGORA project, as summarized below.

Radiative cooling and the ultraviolet background (UVB) are calculated using the cooling library \textsc{Grackle}\footnote{The code's website is \url{https://grackle.readthedocs.io/}} \citep{2017MNRAS.466.2217S}. 
We adopt the UVB model by \citet[][HM12]{2012ApJ...746..125H}, which is activated at $z=15$. Additionally, we apply a floor at the temperature of the cosmic microwave background (CMB).
We also apply the non-thermal Jeans pressure floor to prevent artificial fragmentation \citep{1997ApJ...489L.179T,2016ApJ...833..202K}.
A star particle has a mass of more than $6.1\times10^4\, M_\odot$ at creation and is formed stochastically at a rate of 
\begin{equation}
\dot\rho_* = \epsilon_*\frac{\rho_{\rm gas}}{t_{\rm ff}} 
\end{equation}
when the hydrogen number density exceeds $1\,{\rm cm}^{-3}$, where $\epsilon_*=0.01$ is the star formation efficiency, and $t_{\rm ff} = \sqrt{3\pi/32G\rho_{\rm gas}}$ is the local free-fall time. 

Unlike AGORA's common physics, we use the non-equilibrium cooling mode that calculates the abundances of electrons, \ion{H}{1}, \ion{H}{2}, \ion{He}{1}, \ion{He}{2}, and \ion{He}{3} using \textsc{Grackle v3.3}.
We also use the self-shielding option 3 of \textsc{Grackle}, which considers self-shielding in \ion{H}{1} and \ion{He}{1} using the fitting formula of \citet{2013MNRAS.430.2427R}, but ignores \ion{He}{2} ionization and heating from the UV background.

We evolve the initial condition to $z=5.5$ with three different UVB intensities. We call the run evolved with the default HM12 UVB intensity the \texttt{HM12UVB} run, and the runs with three and one-third times the default intensity \texttt{StrongUVB} and \texttt{WeakUVB} runs, respectively.

We note that stellar feedback, including radiative feedback, is not included, and we assume a uniform UV background radiation as an ionizing source.

\subsection{Results}

\begin{figure}
    \centering
    \includegraphics[width=\linewidth]{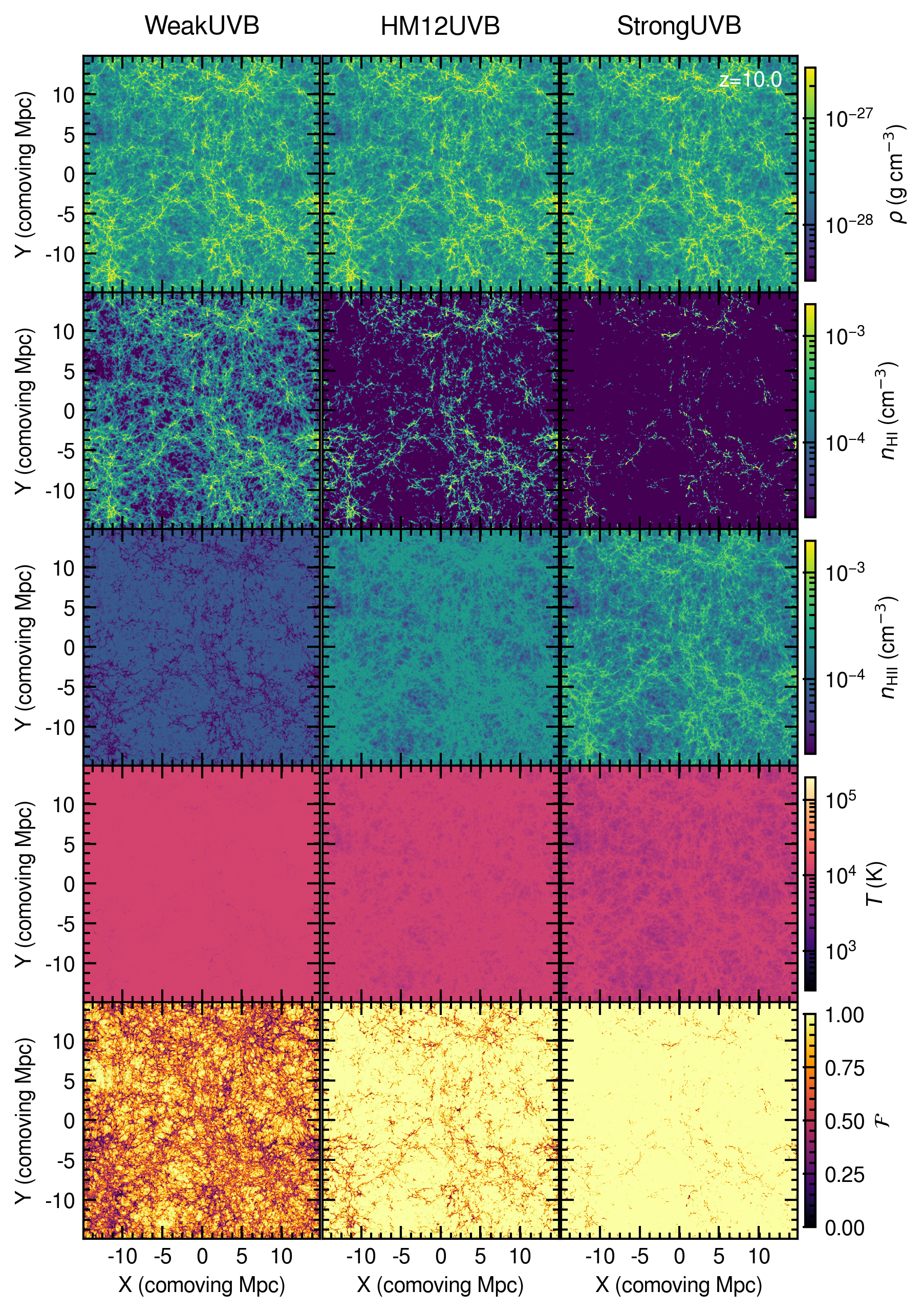}
    \caption{shows slices of the \texttt{WeakUVB} run (left), \texttt{HM12UVB} run (middle), and \texttt{StrongUVB} run (right) at $z=10$. From top to bottom: gas density, \ion{H}{1} number density, \ion{H}{2} number density, temperature, and effective transmission factor of ionizing photon.}
    \label{fig: projections}
\end{figure}

Figure~\ref{fig: projections} shows the slices of the gas density, \ion{H}{1} density, \ion{H}{2} density, temperature, and effective transmission factor at $z=10$.
The \texttt{WeakUVB}, \texttt{HM12UVB}, and \texttt{StrongUVB} runs exhibit nearly identical density, as they evolve from the same initial condition.
The \ion{H}{1} density increases for runs with weaker UVB intensity, leading to a corresponding decrease in the effective transmission factor.
In contrast, the temperature in the IGM is lower for runs with higher UVB intensities. With stronger UVB, hydrogen is ionized earlier, resulting in reduced photoionization heating at $z=10$.

\begin{figure}
    \centering
    \includegraphics[width=\linewidth]{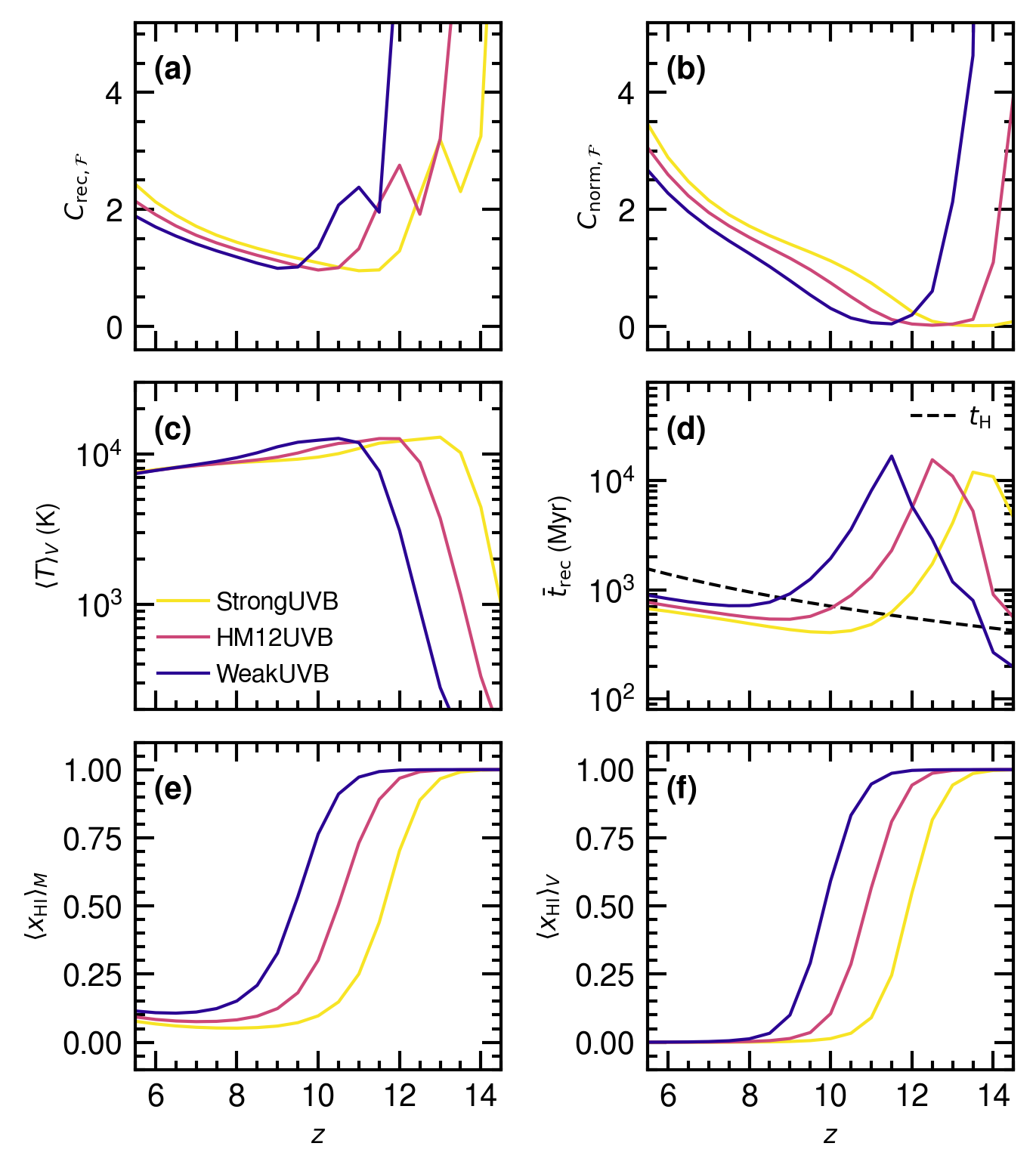}
    \caption{shows redshift evolution of the (a) recombination clumping factor, (b) normalized clumping factor, (c) volume-averaged temperature, (d) mean recombination time, (e) mass-weighted average of the neutral fraction of hydrogen, and (f) volume-weighted average of the neutral fraction of hydrogen.
    The yellow, magenta, and blue lines are from \texttt{StrongUVB}, \texttt{HM12UVB}, and \texttt{WeakUVB} runs, respectively.
    The dashed line in the panel (d) indicates the Hubble time.
    }
    \label{fig:unigrid_summary}
\end{figure}

The top left panel of Figure~\ref{fig:unigrid_summary} shows the redshift evolution of the clumping factors from \texttt{HM12UVB}, \texttt{StrongUVB}, and \texttt{WeakUVB} runs.
The $C_{\rm rec, \mathcal{F}}$ as a function of redshift shows a V-shape, with the increase toward higher and lower redshift due to a larger clumpiness of the $\mathcal{F}$ field and \ion{H}{2} density field, respectively.
Stronger UVB intensity causes earlier reionization, shifting the redshift evolution of $C_{\rm {rec}, \mathcal{F}}$ to a higher redshift.

Normalizing the recombination coefficient and $\mathcal{F}$, we obtain the normalized clumping factor $C_{\rm norm, \mathcal{F}}$ as shown in the top right panel of Figure~\ref{fig:unigrid_summary}.
The increase in $C_{\rm norm, \mathcal{F}}$ towards a higher redshift is due to a high $\langle{ \alpha \rangle}_V$ caused by the low IGM temperature at high redshift.

The redshift evolution of the volume-averaged temperature is shown in the middle left panel of Figure~\ref{fig:unigrid_summary}.
The temperature in the IGM increases due to photoheating, which raises the temperature of the gas to $>10^4$ K before the neutral fraction begins to drop significantly.

The middle right panel of Figure~\ref{fig:unigrid_summary} shows the redshift evolution of the recombination time scale averaged in the IGM,
\begin{equation}
    \bar{t}_{\rm rec} = \frac{\langle n_{\rm HII} \mathcal{F} \rangle_V}{\langle \alpha n_{\rm HII} n_{\rm e} \mathcal{F} \rangle_V}.
    \label{eq:trec}
\end{equation}
The recombination timescale decreases as reionization proceeds. Comparison with the Hubble time (dashed line) indicates that recombination is insignificant in the early phase of reionization, while it becomes important in the later phase.

The bottom panels of Figure~\ref{fig:unigrid_summary} show the mass-weighted and volume-weighted averages of the neutral fraction of hydrogen.
They show a similar trend; however, there are differences in the timing of reionization.
The mass-weighted neutral fraction converges to $\sim 0.1$, while the volume-weighted neutral fraction converges to $\sim 0$ at $z=5.5$, indicating that dense regions remain neutral, while underdense regions are fully ionized.

\begin{figure}
    \centering
    \includegraphics[width=\linewidth]{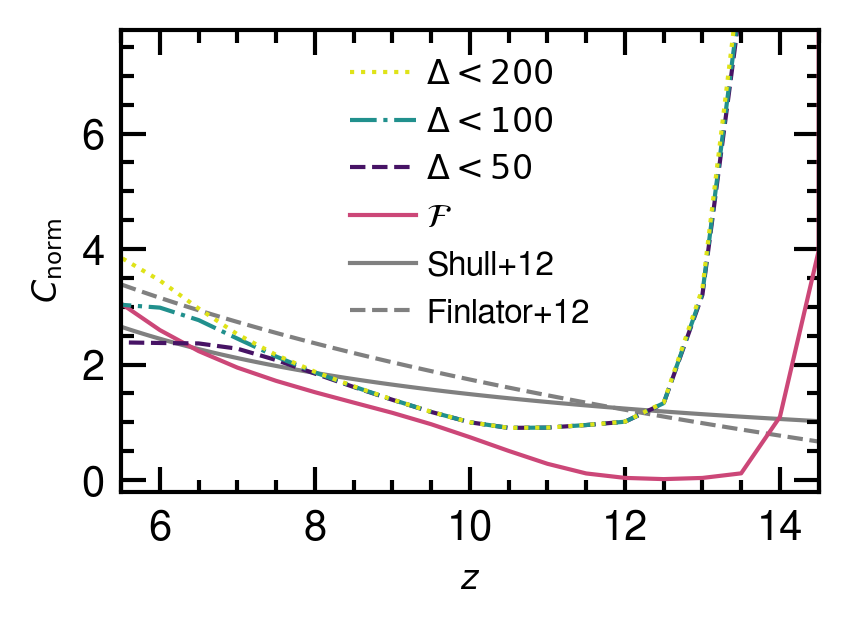}
    \caption{shows redshift evolution of the normalized clumping factor from \texttt{HM12UVB}. The solid line shows the $C_{{\rm norm}, \mathcal{F}}$, and dashed, dot-dashed, and dotted lines are $C_{{\rm norm}, \Delta_{\rm th}}$ with $\Delta_{\rm {th}}=50$, 100, and 200, respectively.
    The gray solid and dashed lines display fitting functions from \citet{2012ApJ...747..100S} and \citet{2012MNRAS.427.2464F}, respectively.
    }
    \label{fig:clumpingfactor_delta}
\end{figure}

Figure~\ref{fig:clumpingfactor_delta} compares the redshift evolution of normalized clumping factors obtained using different IGM criteria from the \texttt{HM12UVB} run.
The solid magenta line is the $C_{\rm {norm},\mathcal{F}}$, as defined in Eq.~(\ref{eq:cnorm}), which is identical to the magenta line in the top right panel of Figure~\ref{fig:unigrid_summary}.
Instead of using $\mathcal{F}$, one can consider the IGM as the region where the gas overdensity is below a threshold and define the clumping factor as
\begin{equation}
    C_{\rm rec, \Delta_{\rm {th}}} = 
    \frac{
    \langle{  \alpha  n_{\rm HII} n_{\rm e} \theta(\Delta_{\rm th} - \Delta)  }\rangle_{V}
    }
    {  
    \langle{ \alpha   }\rangle_{V}
    \langle{ n_{\rm HII}       }\rangle_{V}
    \langle{ n_{\rm e}         }\rangle_{V}
    \langle{ \theta( \Delta_{\rm th} - \Delta) }\rangle_{V}
    },
    \label{eq:crecdelta}
\end{equation}
where $\theta$ is the Heaviside step function.
Then, the normalized form becomes
\begin{equation}
    C_{\rm norm, \Delta_{\rm {th}}} = 
    \frac{
    \langle{  \alpha  n_{\rm HII} n_{\rm e} \theta(\Delta_{\rm th} - \Delta)  }\rangle_{V}
    }
    {  
    \alpha (10^4\, {\rm K})
    \langle{ n_{\rm HII}       }\rangle_{V}
    \langle{ n_{\rm e}         }\rangle_{V}
    }.
    \label{eq:cnormdelta}
\end{equation}
The dashed, dot-dashed, and dotted lines in Figure~\ref{fig:clumpingfactor_delta} are the normalized clumping factors obtained with $\Delta_{\rm th}=50$, 100, and 200, respectively.
The clumping factors obtained with different IGM criteria show similar redshift evolution at $z<10$, although $C_{\rm norm, \mathcal{F}}$ is lower than $C_{\rm norm, \Delta_{\rm {th}}}$ at $z>6.5$ because $\mathcal{F}$ excludes self-shielded cells that tend to have a higher recombination rate.
Differences in the threshold overdensity $\Delta_{\rm th}$ lead to variations in $C_{\rm norm, \Delta_{\rm {th}}}$ at $z<7.5$, with higher thresholds yielding larger clumping factors.
We also compare our results with fitting functions from \citet{2012ApJ...747..100S} and \citet{2012MNRAS.427.2464F}, which are shown as gray solid and dashed lines, respectively.
We find that our results are consistent within a factor of two with these previous studies at $z<10$, when the IGM is predominantly ionized in the \texttt{HM12UVB} run.

\subsection{Discussion}

The similarity of the clumping factor value, $C \sim 3$, in both previous studies and this work indicates that the clumping factor is robust to numerical resolution and box size.
The static-grid simulations by \citet{2012ApJ...747..100S} used a box length of $50\,h^{-1}\,{\rm Mpc}$ with up to $1536^3$ cells, corresponding to a cell size of 32.5 comoving kpc. The SPH simulations by \citet{2012MNRAS.427.2464F} used a box length of $6\,h^{-1}\,{\rm Mpc}$ with $256^3$ particles, corresponding to a baryonic mass resolution of $2.3\times10^5\, M_\odot$.
Our AMR simulation has a box length of $20\,h^{-1}\,{\rm Mpc}$, a root-grid resolution of $1024^3$, and adaptive mesh refinement up to level 4 with a baryonic mass refinement criterion of $1.51\times10^5\, M_\odot$. The finest cell size, 1.81 comoving kpc, resolves the Jeans length of cosmic filaments in the IGM and is sufficient to capture the clumpiness of the diffuse IGM.

We also find that the clumping factor depends only weakly on the IGM definition within the range of definitions commonly adopted in the literature.
At $z<10$, $C_{\rm norm, \mathcal{F}}$ and $C_{\rm norm, \Delta_{\rm th}}$ (with $\Delta_{\rm th}=50$, 100, and 200) show similar redshift evolution.
The agreement between the effective-transmission-based clumping factor and the overdensity-based clumping factor suggests that the typically adopted density threshold of $\Delta_{\rm th} \sim 100$ effectively captures the clumpiness of the diffuse IGM.

\section{Zoom-in Simulations}
\label{sec:zoom}
Following up on the findings from the large-scale simulations in the previous section, we now investigate the impact of stellar feedback on the recombination clumping factor.
We carry out two zoom-in cosmological simulations with and without stellar feedback from an identical initial condition.

\subsection{Method}
We generate the initial condition at $z=99$ using \textsc{Music} with the set of parameters taken from the AGORA project\footnote{The project's website is \url{https://sites.google.com/site/santacruzcomparisonproject/}} \citep{2014ApJS..210...14K,2021ApJ...917...64R} tagged 1e12q.
We assumed a flat $\Lambda$ CDM cosmology with cosmological parameters $(\Omega_{\rm m}, \Omega_{\rm \Lambda}, \Omega_{\rm b}, \sigma_8, n_{\rm s}, h)=(0.272, 0.728, 0.0455, 0.807, 0.961, 0.702)$, consistent with WMAP7/9+SNe+BAO \citep{2011ApJS..192...18K, 2013ApJS..208...19H}.
The simulation box of $(60\,h^{-1}\,{\rm cMpc})^3$ is covered by root grids of $128^3$ with a series of five nested higher-resolution grids; The finest grid level in the initial condition has the equivalent unigrid resolution of $4096^3$.
The zoom-in region covered with the finest grids has an ellipsoidal shape that encloses all the dark matter particles that eventually end up within $4R_{\rm vir}$
of the target halo at $z=0$.
Consequently, at the highest level, the particle mass of dark matter is $m_{\rm DM} = 2.8\times10^5\, M_\odot$, and the mean mass of the gas cell is $m_{\rm gas, IC}=5.65\times10^4\, M_\odot$.

We carry out cosmological hydrodynamical simulations using \textsc{Gamer-2} with the same numerical scheme as in the previous section and AGORA's common subgrid physics model; i.e., we use the equilibrium cooling mode of \textsc{Grackle} with the default HM12 UVB model.
The hydrogen ionization state is obtained using a table generated with \textsc{Cloudy v23.01} \citep{2023RMxAA..59..327C} in post-processing.
The mesh refinement is allowed to reach an additional seven refinement levels; thus, the finest refinement level is 12, and its cell size is 163 comoving pc.

Feedback from type II and Ia supernovae (SNe) and asymptotic giant branch (AGB) stars is modeled as presented in \citet{2022ApJS..262....9O} and \citet{2024ApJ...975..183O}, where the model has been incorporated into the smoothed particle hydrodynamic code \textsc{Gadget4-osaka} and is successfully used in the CROCODILE cosmological simulation\footnote{The project's website is \url{https://sites.google.com/view/crocodilesimulation/home}}.
We assumed a simple stellar population and generate the metal and energy yield tables for SNe and AGB stars using the chemical evolution library \textsc{Celib}\footnote{The code's website is \url{https://bitbucket.org/tsaitoh/celib}} \citep{2017AJ....153...85S} with the following assumptions.
We adopt the Chabrier IMF \citep{2003PASP..115..763C} at solar metallicity and add a log-flat component of massive stars at low metallicities, using the fitting function derived from star cluster formation simulations by \citet{2022MNRAS.514.4639C}.
We use the type II SN yield model by \citet{2013ARA&A..51..457N}, which covers the progenitor mass range of $[13\, M_\odot, 40\, M_\odot]$.
We assume that 30\% of those in the progenitor mass range of $[20\,M_\odot,40\,M_\odot]$ explode as hypernovae (HNe) when the stellar metallicity is $Z<10^{-3}$. The HNe have an explosion energy of $10^{52}\,{\rm erg}$, an order of magnitude larger than normal SNe. The HNe fraction is dropped to 1\% at higher metallicities.
Considering the possible contributions by SNe from lower-mass progenitors and the missing feedback effects from stellar winds and radiation, we boost the type II SN energy by a factor of three.
For type Ia SNe, we assume the power-law-type delay-time distribution \citep{2012PASA...29..447M} and the element yield model by \citet{2013MNRAS.429.1156S}. For AGB stars, we use the element yield table by \citet{2010MNRAS.403.1413K} and \citet{2014MNRAS.437..195D}.

We inject the terminal momentum of SN remnants into the cells surrounding the feedback site, considering the radiative cooling loss on an unresolved scale \citep{2014ApJ...788..121K, 2017ApJ...846..133K, 2018MNRAS.477.1578H}.
We estimate the momentum input from superbubbles formed by clustered SNe using the fitting function of momentum injection per SN, derived from 3D simulations of superbubbles in a variety of density and metallicity environments, as well as different SN explosion intervals by \citet{2022ApJS..262....9O},
\begin{equation}
    \hat p = 1.75\times10^5\,M_\odot\,{\rm km}\,{\rm s}^{-1}\,n_0^{-0.55}\,\Lambda_{6,-22}^{-0.17},
\end{equation}
where $n_0 = n_{\rm H}/(1\,{\rm cm}^{-3})$ is the normalized hydrogen number density, and $\Lambda_{6,-22}^{-0.17}=\max\{1.9-0.85(Z/0.0194),1.05\}\times(Z/0.0194)+10^{-1.33}$ is the value of the cooling function of \citet{1993ApJS...88..253S} at $T=10^6\,{\rm K}$, normalized by $10^{-22}\,{\rm erg}\,{\rm s}^{-1}\,{\rm cm}^3$.
We enforce energy conservation when the increase in kinetic energy due to the momentum input exceeds the SN energy.

We perform two zoom-in simulations with and without stellar feedback, which correspond to the \texttt{Cal-3} and \texttt{Cal-4} in the calibration steps of the AGORA \texttt{CosmoRun}.
We name the run without feedback the \texttt{NoFB} run and the run with SN feedback the \texttt{SNFB} run.

\subsection{Results}

\begin{figure}
    \centering
    \includegraphics[width=\linewidth]{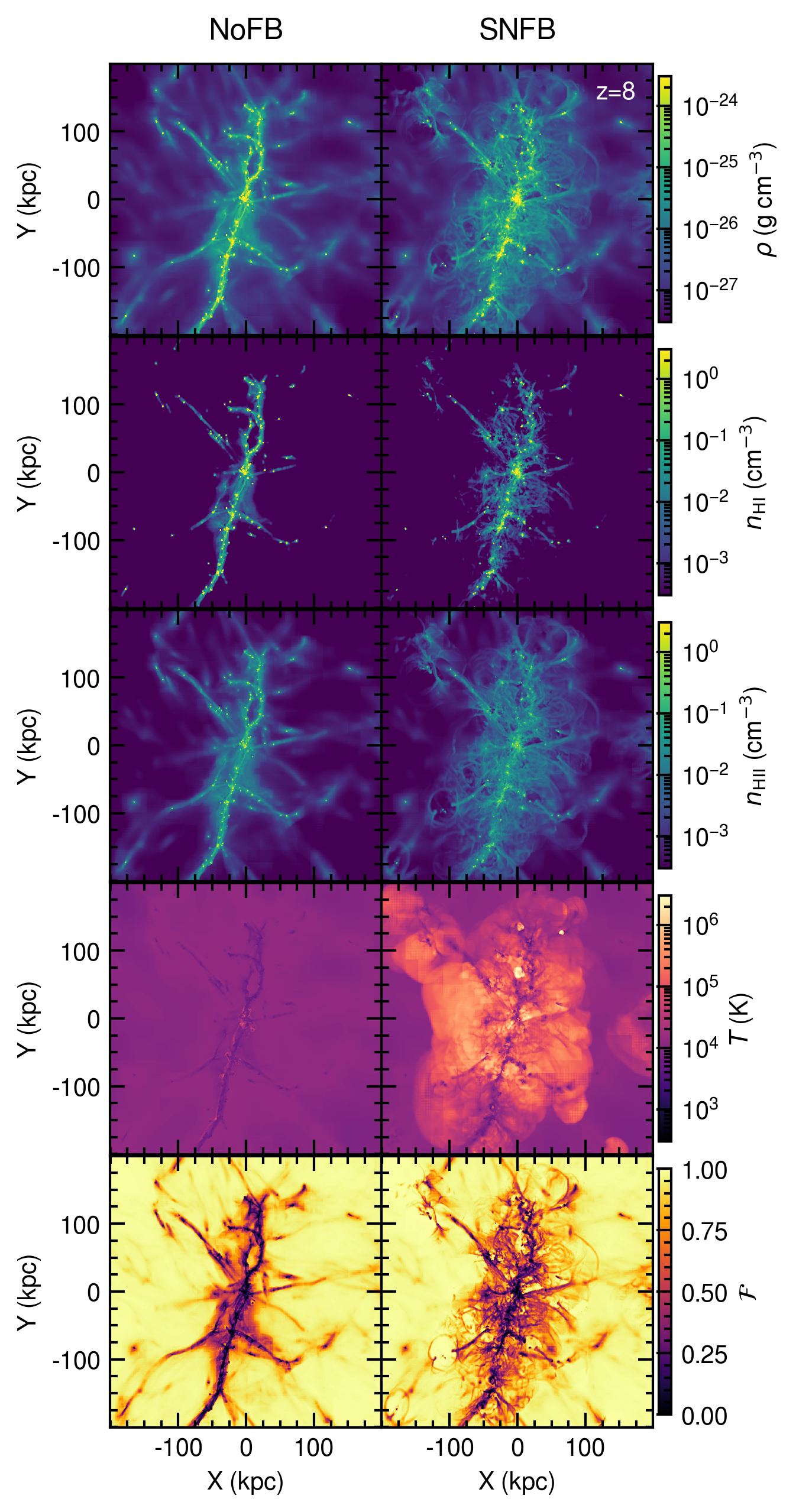}
    \caption{shows density-weighted projections from \texttt{NoFB} run (left) and \texttt{SNFB} run (right) with projection width and depth of $400\, {\rm kpc}$ (physical) at $z=8$. From top to bottom: gas density, \ion{H}{1} number density, \ion{H}{2} number density, temperature, and effective transmission factor of ionizing photon.}
    \label{fig:zoom_projection}
\end{figure}

Figure~\ref{fig:zoom_projection} shows the projections of density, \ion{H}{1} number density, \ion{H}{2} number density, temperature, and the effective transmission factor from the \texttt{NoFB} and \texttt{SNFB} runs at $z=8$.
The dark matter halo mass of the central galaxy is $1.1\times10^{10}\,M_\odot$ at this redshift in both runs.
The stellar mass of the central galaxy at this redshift is $M_* = 2.34\times10^9\,M_\odot$ and $M_* = 7.84\times10^6\,M_\odot$ in the \texttt{NoFB} and \texttt{SNFB} runs, respectively.
The feedback effect is most visible in the comparison of the temperature projections between \texttt{NoFB} and \texttt{SNFB} runs.
The high-temperature bubbles ($T\gtrsim10^5\, {\rm K}$) formed by SN feedback are apparent in the \texttt{SNFB} run, which covers $\sim100\, {\rm kpc}$ around the low-temperature ($T\sim10^3\, {\rm K}$) cosmic filament.
The IGM unaffected by feedback has the equilibrium temperature, $T\sim10^4\, {\rm K}$. 
The bubbles are also identified in the density projection in the \texttt{SNFB} run, while the \texttt{NoFB} run shows a smooth density structure in the IGM.
The ionized hydrogen follows the gas distribution, except for the high-density peaks where hydrogen becomes neutral.
The ionized hydrogen in the \texttt{SNFB} run spreads to a larger distance from the central galaxy compared to the \texttt{NoFB} run.
The neutral hydrogen is distributed along the cosmic filament, where the ionizing photons are self-shielded ($\mathcal{F} \sim 0$).

\begin{figure}
    \centering
    \includegraphics[width=0.7\linewidth]{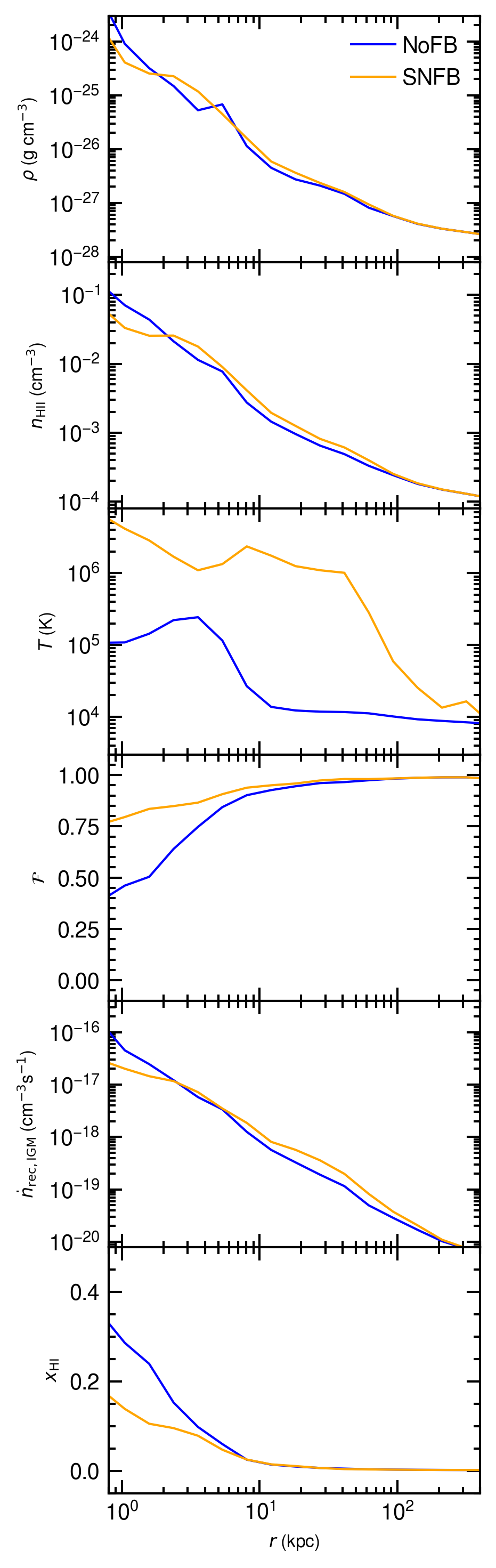}
    \caption{shows spherically averaged quantities as a function of radius in physical kpc around the central galaxy from \texttt{NoFB} run (blue) and \texttt{SNFB} run (orange) at $z=8$. From top to bottom: density, \ion{H}{2} number density, temperature, effective transmission factor, recombination rate, and neutral fraction.}
    \label{fig:zoom_radprof}
\end{figure}

Figure~\ref{fig:zoom_radprof} shows the radial profiles of density, \ion{H}{2} number density, temperature, effective transmission factor, recombination rate, and neutral fraction around the central galaxy at $z=8$.
The \texttt{SNFB} run shows a higher density and temperature than the \texttt{NoFB} run at $2\, {\rm kpc} < r < 100\, {\rm kpc}$ due to the galactic outflow driven by SN feedback.
The SN feedback also smooths out the neutral hydrogen and increases the $\mathcal{F}$ by $\lesssim 50\%$ at $r < 100\, {\rm kpc}$.
The recombination rate, given by $\dot{n}_{\rm rec, IGM} = \alpha n_{\rm HII} n_{\rm e} \mathcal{F}$, is higher in the region $2\, {\rm kpc} < r < 100\, {\rm kpc}$ for the \texttt{SNFB} run due to the higher $n_{\rm {HII}}$ and $\mathcal{F}$, even though the higher temperature reduces the recombination coefficient. Nonetheless, SN feedback contributes to ionization by heating, thereby reducing the neutral fraction near the center.

\begin{figure}
    \centering
    \includegraphics[width=0.8\linewidth]{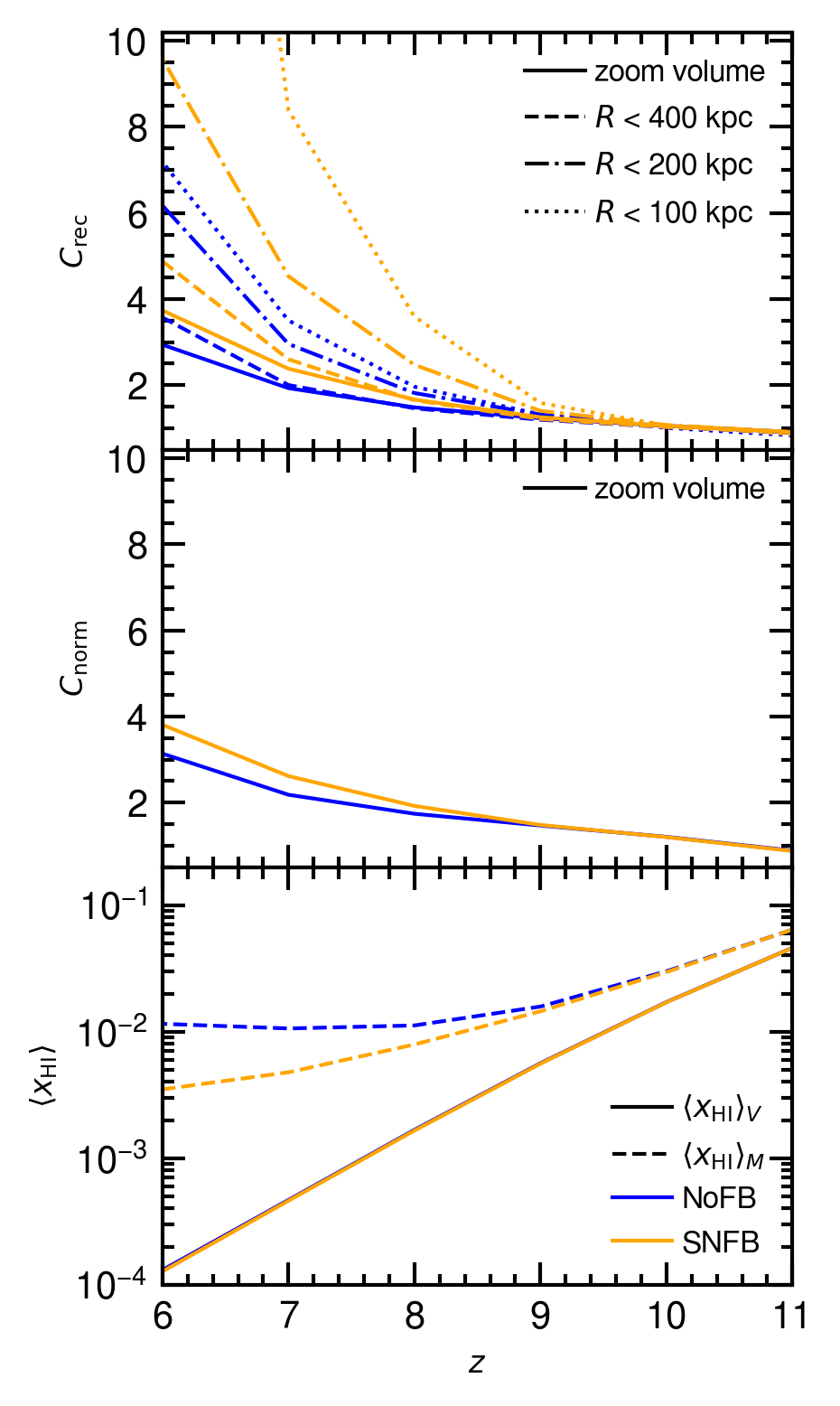}
    \caption{compares the redshift evolutions of clumping factor and neutral fraction from \texttt{NoFB} (blue) and \texttt{SNFB} (orange) runs.
    Top: The redshift evolution of the recombination clumping factor measured in the zoom volume (solid) and within $400\, {\rm kpc}$ (dashed), $200\, {\rm kpc}$ (dash dotted), and $100\, {\rm kpc}$ (dotted) in physical units around the central galaxy.
    Middle: The normalized clumping factor measured in zoom volume.
    Bottom: The volume-weighted neutral fraction (solid) and mass-weighted neutral fraction (dashed) measured in zoom volume.
    }
    \label{fig:zoom_crec}
\end{figure}

The larger recombination rate in the \texttt{SNFB} run results in a larger clumping factor on a scale of $R<100\, {\rm {kpc}}$, as shown in the top panel of Figure~\ref{fig:zoom_crec}.
The deviation of the \texttt{SNFB} run from the \texttt{NoFB} run increases as the feedback effect becomes stronger with the growth of the central galaxy.
However, the feedback effect is limited to a physical scale of $\sim100\, {\rm {kpc}}$, and the redshift evolution of the recombination clumping factors from \texttt{SNFB} and \texttt{NoFB} are comparable on a scale of $\gtrsim 400\, {\rm {kpc}}$.
The middle panel of Figure~\ref{fig:zoom_crec} shows the normalized clumping factors measured in the whole zoom region from \texttt{SNFB} and \texttt{NoFB} runs, and they also show a comparable redshift evolution, with the difference within $\sim 30\%$.
The bottom panel of Figure~\ref{fig:zoom_crec} compares the redshift evolution of volume-weighted and mass-weighted neutral fractions in the zoom volume from \texttt{SNFB} and \texttt{NoFB} runs, which are also similar to each other.
The ionization effect by SN feedback appears in $\langle x_{\rm HI} \rangle_M$ at $z < 9$, while the $\langle x_{\rm HI} \rangle_V$ shows almost identical evolution between the two runs, which also indicates that the feedback effect is limited to a small volume around the galaxy.

\subsection{Discussion}

We find that the SN feedback has two competing effects on reionization, i.e., enhancing recombination by increasing the density of ionized gas, while facilitating ionization by heating the gas.
Our finding of the SN feedback effect of boosting the clumping factor is consistent with \citet{2012MNRAS.427.2464F}, who reported that galactic winds increase the clumping factor by less than 30\% using cosmological radiation smoothed particle hydrodynamics simulations.
Our analysis further reveals that the spatial scale of the feedback effect is limited to $\sim 100\, {\rm kpc}$ around the galaxy, using high-resolution zoom-in AMR simulations.

We also find that SN feedback increases the transmission factor $\mathcal{F}$, suggesting that feedback can help ionizing photons escape from galaxies, which is consistent with previous simulation studies \citep[e.g.,][]{2014ApJ...788..121K, 2017MNRAS.470..224T,2020ApJ...889L..22C, 2020MNRAS.498.2001M}.
Recent observational studies suggest that radiation feedback from young stars before SN explosions plays a more important role in creating low-density channels for ionizing photons to escape from galaxies \citep{2024A&A...688A.198B,2025ApJ...985..128F,2025ApJ...982..137C}.
Our results do not conflict with these studies, and we find that SN-driven outflows form opaque shells, as can be seen in Figure~\ref{fig:zoom_projection}, which can block ionizing photons from escaping to the IGM.
From a comparison of our controlled simulations with and without SN feedback, we suggest that the SN feedback prepares a favorable condition for radiation feedback, facilitating photon escape by reducing the gas density and neutral fraction in the ISM and inner CGM.

\section{Conclusion}
\label{sec:conclusion}

We have investigated the recombination clumping factor of the diffuse IGM during cosmic reionization using a set of cosmological hydrodynamical simulations powered by the AMR code \textsc{Gamer-2} with a uniform UV background radiation.
Our definition of the clumping factor is derived from the mass-weighted reionization model with an IGM criterion based on the effective transmission factor of ionizing photons.
We have performed three large-scale simulations with different UVB intensities and examined the clumping factor as a function of redshift.
We have also performed two zoom-in simulations with and without SN feedback to study the impact of stellar feedback on the clumping factor.
Our main findings are summarized as follows:

\begin{enumerate}
    \item The recombination clumping factor, based on our IGM definition utilizing the effective transmission factor, $C_{{\rm norm}, \mathcal{F}}$, exhibits comparable redshift evolution to that defined using an overdensity threshold, $C_{\rm norm, \Delta_{\rm th}}$, and aligns with previous studies at $z<10$ in the \texttt{HM12UVB} run, wherein the IGM is predominantly ionized (see Figure~\ref{fig:clumpingfactor_delta}).
    This agreement suggests that the commonly used clumping factor of $C \sim 3$ is robust against the boxsize and the numerical resolution, and the commonly adopted clumping factor defined with an overdensity threshold of $\Delta_{\rm th} = 100$ can be a reasonable approximation for the recombination clumping factor in the late stages of reionization when the IGM is mostly ionized.
    
    \item The SN feedback has two competing effects on reionization, i.e., enhancing recombination by increasing the density of ionized gas, while facilitating ionization by heating the gas and reducing the neutral fraction. 
    The feedback effect is limited to a physical scale of $\sim 100\, {\rm kpc}$ in our simulated galaxy (see Figure~\ref{fig:zoom_radprof}), and the redshift evolution of the recombination clumping factors from simulations with and without SN feedback is comparable on a scale of $\gtrsim 400\, {\rm kpc}$ (see Figure~\ref{fig:zoom_crec}).
    In comparison between zoom-in simulations with and without SN feedback, we suggest that the SN feedback prepares a favorable condition for radiation feedback, facilitating photon escape by reducing the gas density and neutral fraction in the ISM and inner CGM.
\end{enumerate}

Our AMR simulations offer high resolution and improve the estimate of clumpiness in the IGM, but they are limited by the assumption of a uniform UV background radiation.
Cosmological radiation hydrodynamical simulations, along with further analysis using a physical definition of the diffuse IGM, are essential for providing accurate models of the recombination clumping factor.
Further investigations using simulations with different reionization scenarios are necessary to fully understand their impact on the reionization history.
Observations that can probe the reionization topology, such as the angular correlation function of Ly$\alpha$ emitters, the 21\,cm power spectrum, and the Ly$\alpha$ damping wing, provide valuable constraints on these models.

\begin{acknowledgments}
We are grateful to the anonymous referee for thoughtful and constructive comments, whose careful and sustained engagement has significantly improved this manuscript.
This work is supported in part by the National Key Research and Development Program of China and the Zhejiang Provincial Top-Level Research Support Program.
The analysis presented in this article was carried out on the SilkRiver Supercomputer of Zhejiang University. 
\end{acknowledgments}

\bibliography{ref}{}
\bibliographystyle{aasjournalv7}

\end{document}